\documentclass[runningheads]{llncs}

\usepackage{graphicx}
\usepackage{lscape}

\begin{document}

\title{Scalability Model Based on the Concept of Granularity}

\titlerunning{Scalability Model Based on the Concept of Granularity}

\author{Jan Kwiatkowski\inst{1} \and \L{}ukasz P. Olech\inst{2}}

\authorrunning{J. Kwiatkowski and \L{}. P. Olech}

\pagestyle{empty}

\institute {Department of Informatics, Faculty of Computer Science and Management, Wroclaw University of Technology, Wybrzeze Wyspianskiego 27, 50-370 Wroclaw, Poland\\
	\email{jan.kwiatkowski@pwr.edu.pl}
    \and
Department of Artificial Inteligence, Faculty of Computer Science and Management, Wroclaw University of Technology, Wybrzeze Wyspianskiego 27, 50-370 Wroclaw, Poland \\
	\email{lukasz.olech@pwr.edu.pl}}

\maketitle

\begin{abstract}
In the recent years it can be observed increasing popularity of parallel processing using multi-core processors, local clusters, GPU and others. Moreover, currently one of the main requirements  the IT users is the reduction of maintaining cost of the computer infrastructure. It causes that the performance evaluation of the parallel applications becomes one of the most important problem. Then obtained  results allows efficient use of available resources. In traditional methods of performance evaluation the results are based on wall-clock time measurements. This approach requires consecutive application executions and includes a time-consuming data analysis. In the paper an alternative approach is proposed. The decomposition of parallel application execution time onto computation time and overheads related to parallel execution is use to calculate the granularity of application and then determine its efficiency. Finally the application scalability can be evaluates.

\end{abstract}

\section{Introduction}

In the recent years there has been rapid development of new technologies related to the evolution of the technical possibilities offered by computer hardware - increasing calculation speed, decreasing communication time, increasing bandwidth communications, etc. Moreover it can be observed increasing popularity of parallel processing by using multi-core processors, clusters, GPU and others. Equally important as the evolution of the information systems are changes of the requirements of the IT users. Increasingly, the basic requirement of the IT users are not systems, offering improved processing speed, but ones that will reduce the cost of maintaining infrastructure. It causes that performance evaluation constitutes an intrinsic part of every application development process. In parallel programming the goal of the design process is not to optimise a single metrics, a good design has to take into consideration memory requirements, communication cost, efficiency, implementation cost, and others. Therefore performance evaluation of parallel programs is very important for the development of efficient parallel applications. 

In the paper \cite{jogalekar} three categories of performance metrics have been proposed. The first are speedup metrics that show how faster results can be obtain when using some number of processing units comparing with using only one processing unit. The second one are efficiency metrics that determine the percentage of  CPU utilization during parallel program execution. And finally scalability, which say how application behaves when increasing the number of available processing units and/or the size of the problem being solved. In the paper all of these metrics will be used for performance evaluation of parallel application.
  
In general the performance analysis can be carried out analytically or through experiments. The paper focusses on the second approach. Independently on the used measurement method during experimental performance evaluation of parallel programs is the need to measure the run time of sequential and parallel programs, which is time consuming. In the paper the method, which overcomes above problem is proposed. Basing on the concept of granularity and decomposition of the parallel application execution time onto the computation time and the overhead time presented in \cite{kwiatkowski_2,kwiatkowski_3} we show that by measurement only wall-clock time and computation time it is possible to evaluate the performance of parallel programs. The paper extends previous one by presentations results of experiments performed up to 4096 processing units (cores) and by scalability analysis.

The paper is organised as follows. Section 2 briefly describes
different performance metrics and two main approaches to scalability analysis - strong and weak scalability. How  granularity can be used in performance evaluation is presented in section 3. The next section illustrates the experimental results obtained during evaluation of two parallel algorithms, strong and weak scalability are considered. Finally, section 5 outlines the work and discusses ongoing work. 

\section{Performance Metrics and scalability analysis}

During performance evaluation of parallel applications different metrics are used \cite{grama}. The first one is the parallel run time $(t_{runtime})$. It is the time from the moment when computation starts to
the moment when the last processor finishes its execution and is composed of three different times: computation time $(t_{comp})$ is the time spent on performing computation by all processors, communication time $(t_{comm})$ is the time spent on sending and receiving messages by all processors and idle time $(t_{idle})$ is when processors stay idle. The next commonly used metric is speedup, which captures the relative benefit of solving a given problem using a parallel system. There exist different speedup definitions. Generally the speedup $(S)$ is defined as the ratio of the time needed to solve the problem on a single processor to the time required to solve the same problem on a parallel system with $p$ processors. Theoretically, speedup
cannot exceed the number of processors used during program execution, however, different speedup anomalies can be observed \cite{kwiatkowski_1}. Both above mentioned performance metrics do not take into account the utilisation of processors in the parallel system. While executing a parallel algorithm processors spend some time on communicating and some processors can be idle. Then the efficiency $(E)$ of a parallel program is defined as a ratio of speedup to the number of used processors. In the ideal parallel system the efficiency
is equal to one but in practice efficiency is between zero and
one, however because of different speedup anomalies, it can be even 
greater than one. 

The last performance metrics is scalability of the parallel system. It can be considered in different ways, we can  use it for hardware, algorithms, data bases, execution environment, etc.  One can say that currently it is one of the most important performance metrics. In general it can be say that it is a metrics, which consider the "system" capacity to increase speedup in proportion to the number of available processors. There are a lot of approaches to modelling the scalability, for example by using so called isoefficiency analysis \cite{grama}, Universal Scalability Model proposed by the Neil Gunther \cite{gunther}, H-isoefficiency function \cite{bosque} and others. 

One can find two different approaches to way in which scalability is defined \cite{shoukourian}. The first one based on Amdahl law (\ref{eq1}) is called strong scalability. The strong scalability is also called scalability with a fixed size of the problem, it means that our goal is to minimize the program execution time by using more processing units. It means that we can say that system is scalable when increasing number of processing units are used effectively. For example, when the number of processing units equals 8 and the speedup received equals 8, too, then we have excellent scalability. This approach is the pessimist because of indicates a bounded speedup. 

\begin{equation}
\label{eq1} 
\mathcal Speedup(n) = \frac{T(1)}{T(n)} =
\frac{1}{(1-p) + \frac{p}{n}} 
\end{equation}

\noindent where $n$ denotes the number of processing units, $p$ denotes the non-scaled fraction of the application parallel part and $T(1)$, $T(n)$ execution time at $1$ and $n$ processors respectively. 

The second one is weak scalability that based on Gustafson law  (\ref{eq2}). The week scalability is also called the scalability with variable problem size, when the problem size increased at the time when the number of processing units increased (the input is fixed for each processor). We say that a system is scalable when the efficiency (execution time)is the same for increasing the number of processors and the size of the problem \cite{grama}. This approach is the optimistic because of indicates an unlimited speedup.

\begin{equation}
\label{eq2} 
\mathcal Speedup(n) = \frac{T(1)}{T(n)} =
1 + (n-1)*p^*
\end{equation}

\noindent where $n$ denotes number processing units, $p^*$ denotes the scaled fraction of the application parallel part and $T(1)$, $T(n)$ execution time at $1$ and $n$ processors respectively. 

\section{Using Granularity for Performance Analysis}

In general the granularity of a parallel computer is defined as a ratio of the time required for a basic communication operation to the time required for a basic computation operation. Let's define the granularity of the parallel algorithm similarly as the ratio of the amount of computation to the amount of communication within a parallel algorithm execution $(G=T_{comp}/T_{comm})$. Above definition can be used for calculating the granularity of a single process executed during program execution on each processor as well as
for the whole program by using total communication and computation times of all program processes. 
Then let's use the overhead function, which is a function of problem size and the number
of processors and is defined as follows \cite{grama}:

\begin{equation}
\label{eq3}
\mathcal T_o (W,p) = p\ast T_p - W
\end{equation}

\noindent where $W$ denotes the problem size, $T_{p }$ denotes time
of parallel program execution and $p$ is the number of used processors.

The problem size can be defined as the number of basic computation operations required to solve the problem using the best serial algorithm. Let us assume that a basic computation operation takes one unit of time. Thus the problem size is equal to the time of performing the best serial algorithm on a serial computer. Based
on the above assumptions after rewriting the equation (\ref{eq3}) we obtain the following expression for parallel run time:

\begin{equation}
\label{eq4}
\mathcal T_p = \frac{W + T_o (W,p)}{p}
\end{equation}

Recalling that the parallel run time consists of computation time, communication time and idle time, let's assume that the main overhead of parallel program execution is communication time.
The total communication time is equal to the sum of the communication time of all performed communication steps. Assuming that the distribution of data among processors is equal then the communication time can be calculated using equation $T_{total\_comm }= p * T_{comm }$. Note that the above is true
when the distribution of work between processors and their
performance is equal. Similarly, the computation time is the sum
of the time spent by all processors performing computation. Then
the problem size $W$ is equal to $p* T_{comp}$. Therefore the expression for the efficiency takes the form: 

\begin{equation}
\label{eq5} 
\mathcal E = \frac{1}{1 + \frac{T_{comm} }{T_{comp} }} =
\frac{1}{1 + \frac{1}{G}} = \frac{G}{G + 1}
\end{equation}

It means that using the concept of granularity we can calculate the efficiency
and speedup of parallel algorithms. Concluding above consideration it is possible to evaluate a parallel application using such metrics as efficiency, speedup and scalability by measuring only the computation and wall-clock times during execution of parallel version of a program on a parallel computer. Deeper presentation of the above discussion can be find in \cite{kwiatkowski_2}.

\section{Case studies}
To confirm the usefulness of the theoretical analysis presented in the previous
sections the series of experiments were performed. During the experiments two different algorithms were
used: K-means and Monte Carlo method (calculation of Pi number). The tests were executed on the BEM cluster at Wroclaw Centre for Networking and Supercomputing (720 homogeneous nodes (2 procesors) Intel Xeon E5 – 2670 v3).  
For both algorithms the strong scalability was checked and weak scalability was check only for K-means algorithm.

To avoid the execution time anomalies \cite{kwiatkowski_2} the experiments were performed for data sizes
sufficiently larger than CPU cache size and smaller than the main memory limits for strong scalability analysis and for weak scalability analysis the problem size increased proportionally to the number of used processors. Because the experiments were performed in a multi-user environment the execution times depended on computer load, therefore the presented results are the averages from the series of 10 identical experiments performed. Moreover the results of measurement lying in the distance above 1.5 interquartile range of the whole series were treated as erroneous and omitted, and the measurement was repeated. To evaluate the accuracy of the new method the relative error defined as $\frac{\mathcal{S} - S}{S}$ where $S$ is the actual speedup and $\mathcal{S}$ is the estimated one has been used. Moreover because of way in which different times have been measured for speedup calculation instead of granularity isogranularity defined as $(G_{iso}=t_{comp}/t_{overhead})$ was used. 

K-means is one of the algorithms that is used for solving the clustering problem \cite{macqueen}. It classifies a given data set into defined fixed number of clusters $k$ (predefined). In the first algorithm's step so called centroids for each cluster should be chosen - one for each cluster. These centroids can be defined in random way however the better choice is to place them as much as possible far away from each other. In the next step all points from the data set are assign to the nearest centroid. After completion of this step the new centroids for each cluster are calculated using the means metrics for the created clusters. Then we repeated the second step using these new centroids. The process is continue as long as the differences between coordinates of new and old centroids are satisfied. Alternatively the process can be finished after predefined number of iterations. 

The above algorithm was parallelized in the following simple way. The chosen processor reads input data, and then distributes them to other processors. Each processor received  $N/p$ data, when $p$ is a number of available processors and N is the number of input data. Then each processor generates the appropriate number of centroids and exchanges information about them with other processors. After completion of above step each processor has information about all the centroids and performs the second step of the sequential algorithm. In the next step each processor calculates the data necessary to calculate new centroids (the number of point in each cluster and sums of points coordinates) and exchange this information with other processors. Then the new centroids are calculated, in parallel by all processors  (execution replication), and again the process returns to the second step of sequential algorithm. The algorithm ends when the stop criterion is met. Then the chosen processor collects clustering results from other processors and merge them. Description of Monte Carlo method is skiped because of the common knowledge about it and the lack of space. 

\begin{figure}[bht!]
  \centering
  \includegraphics[scale=.55]{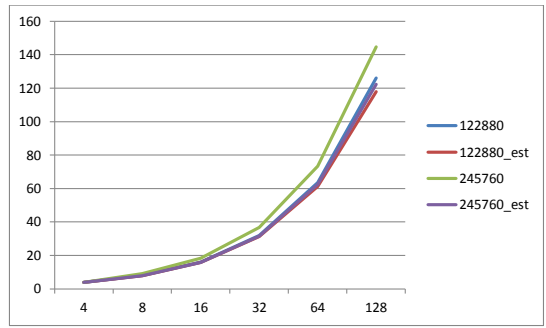}
  \caption{K-means algorithm speedup and estimated speedup - 2 cores at each node}
  \label{fig_k-means_speedup_2_cores}
\end{figure}

\begin{figure}[bht!]
  \centering
  \includegraphics[scale=.55]{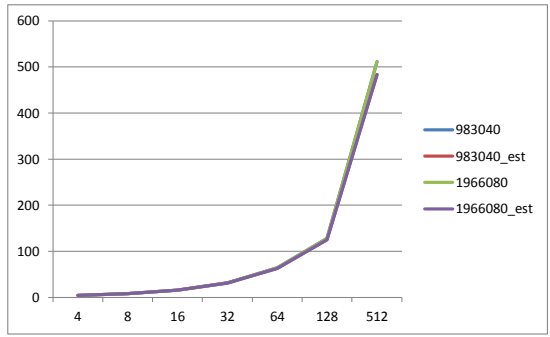}
  \caption{K-means algorithm speedup and estimated speedup - 4 cores at each node}
  \label{fig_k-means_speedup_4_cores}
\end{figure}

\begin{figure}[bht!]
  \centering
  \includegraphics[scale=.55]{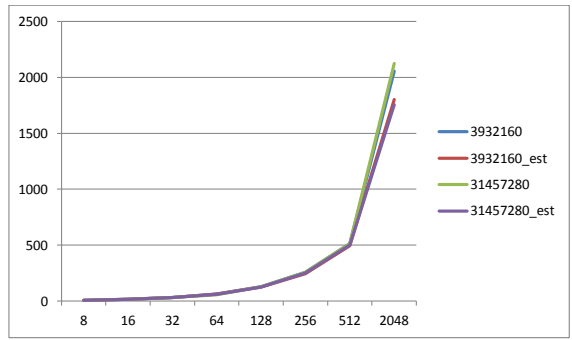}
  \caption{K-means algorithm speedup and estimated speedup - 8 cores at each node}
  \label{fig_k-means_speedup_8_cores}
\end{figure}

\begin{figure}[bht!]
  \centering
  \includegraphics[scale=.55]{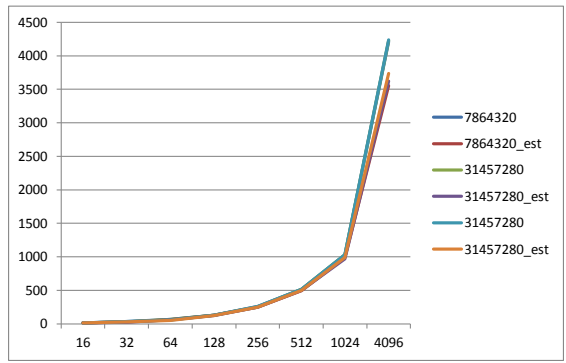}
  \caption{K-means algorithm speedup and estimated speedup - 16 cores at each node}
  \label{fig_k-means_speedup_16_cores}
\end{figure}

Below the results of experiments performed to check the strong scalability of k-means algorithm and Monte Carlo methods are presented. In the experiments performed for K-means algorithm different data set sizes were used. The number of generated clusters was 1024 during all tests. Moreover different hardware configurations by means different number of cores from each processor were used. Received results are presented on figures  \ref{fig_k-means_speedup_2_cores}, \ref{fig_k-means_speedup_4_cores}, \ref{fig_k-means_speedup_8_cores}, \ref{fig_k-means_speedup_16_cores}.

The first test was performed using 2 cores from 2, 4, 8, 16, 32 and 64 processors, its results are presented on figure \ref{fig_k-means_speedup_2_cores}. As can be seen the actual speedup and estimated speedup are very close, however when the size of data set is equal 245760 there are large differences between actual and estimated speedup and the precise relative error is even over 16\% when using 128 cores, for other cases is less than 5\%. 

In the second test 4, 8, 16, 32, 64, 128 and 512 processing units (cores), four from each processor were used. Results of this test are presented on figure \ref{fig_k-means_speedup_4_cores}. As previously can be seen that the actual speedup and estimated speedup are very close. In general the precise relative error was less then 2\%, however for problem size 1966080 was slightly larger when using 512 cores.  

In the third test 8, 16, 32, 64, 128, 256, 512 and 2048 processing units (cores), eight from each processor were used. Results of this test are presented on figure \ref{fig_k-means_speedup_8_cores}. In this test results are really satisfied, the precise relative error were between 1\% and 2\%.  

In the last test performed for K-means algorithm 16, 32, 64, 128, 256, 512, 1024 and 4096 processing units (cores), sixteen from each processor were used. Results of this test are presented on figure \ref{fig_k-means_speedup_16_cores}. As during the previous tests the results were very good, the precise relative error values were between 0,2\% and 6\%, only for problem size equals 7864320 for 4096 processing unites was larger, close to 15\%.

\begin{figure}[bht!]
  \centering
  \includegraphics[scale=.60]{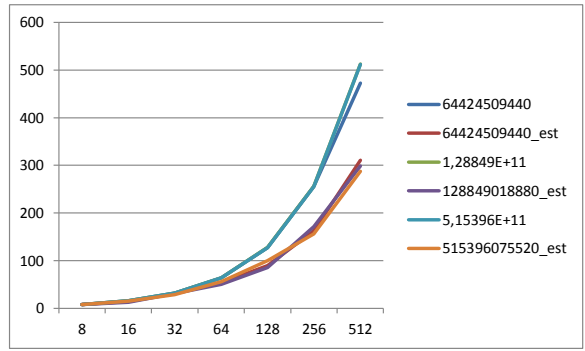}
  \caption{Monte Carlo algorithm speedup and estimated speedup - 8 cores at each node}
  \label{fig_monte_carlo_speedup_8_cores}
\end{figure}

\begin{figure}[bht!]
  \centering
  \includegraphics[scale=.60]{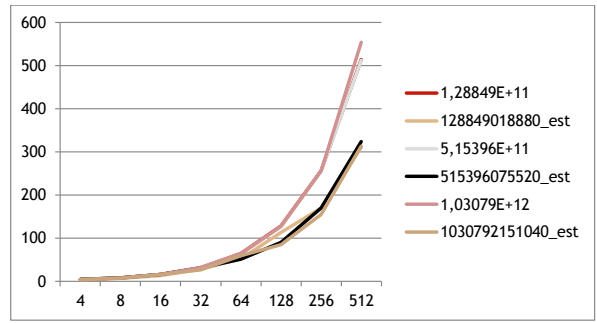}
  \caption{Monte Carlo algorithm speedup and estimated speedup - 16 cores at each node}
  \label{fig_monte_carlo_speedup_16_cores}
\end{figure}

In the experiments performed for Monte Carlo method different data set sizes were used. Moreover different hardware configurations by means different number of cores from each processor were used. Received results are presented on figures  \ref{fig_monte_carlo_speedup_8_cores}, \ref{fig_monte_carlo_speedup_16_cores}.

Results obtained for Monte Carlo methods similarly as for k-means algorithm are very promising, the shape of diagrams are very close and the precise relative error was not larger than 5\% in all cases. Considering the strong scalability we can conclude that results of experiments show that both algorithms are scalable in the limits of defined by the limits of performed tests. 

\subsection{Experimental results - weak scalability}

During test related to cheking weak scalability for k-means algorithm  8, 16, 32, 64, 128, 256, and 512 processing units (cores) randomly chosen have been used. Problem sizes were from 122880 to 31457280 to satisfy requirements that during program execution each processing unit should used the same amount of data. Typically weak scalability is presented as a diagram using scaling efficiency. In the paper we present it in different way proposed in the paper \cite{kartawidjaja} by presenting execution time and speedup in the tables (Table \ref{Tab:tab1}, Table \ref{Tab:tab2}). 

\begin{table}
\caption{Execution time of parallel k-means algorithm}\label{Tab:tab1}
\centering
\begin{tabular}{|c|c|c|c|c|c|c|c|c|}
\hline
Problem size & $T_1$ & $T_8$ & $T_{16}$ & $T_{32}$ & $T_{64}$ & $T_{128}$ & $T_{256}$ & $T_{512}$\\
\hline
122880 & 2,918 & 0,364 & 0,182 & 0,091 & 0,046 & 0,023 & 0,0064 & 0,0047 \\
\hline
983040 & 178,132 & 22,240 & 11,120 & 5,659 & 2,779 & 1,388 & 0,698 & 0,345 \\
\hline
1966080 & 708,634 & 88,522 & 44,262 & 23,137 & 11,063 & 5,5308 & 2,554 & 1,385 \\
\hline
3932160 & 2831,905 & 318,146 & 176,702 & 88,353 & 44,175 & 22,079 & 10,240 & 5,517 \\
\hline
7864320 & 11337,77 & 1416,032 & 812,677 & 353,901 & 176,960 & 88,462 & 38,918 & 21,949 \\
\hline
15728640 & 45318,33 & 5660,593 & 2830,199 & 1624,234 & 707,452 & 353,600 & 176,745 & 88,342 \\
\hline
31457280 & 181417,6 & 22653,32 & 11325,1 & 5661,966 & 3249,23 & 1414,826 & 706,802 & 353,248 \\
\hline
\end{tabular}
\end{table}

\begin{table}
\caption{Speedup of parallel k-means algorithm based on Gustafson's model}\label{Tab:tab2}
\centering
\begin{tabular}{|c|c|c|c|c|c|c|c|}
\hline
Problem size & $S_8$ & $S_{16}$ & $S_{32}$ & $S_{64}$ & $S_{128}$ & $S_{256}$ & $S_{512}$\\
\hline
122880 & 7,946 & 15,744 & 31,297 & 61,302 & 117,931 & 70,688 & 0,0047 \\
\hline
983040 & 7,948 & 15,822 & 31,431 & 62,546 & 122,945 & 240,221 & 408,803 \\
\hline
1966080 & 7,972 & 15,868 & 31,232 & 62,909 & 125,105 & 81,220 & 483,412 \\
\hline
3932160 & 7,958 & 15,843 & 31,630 & 63,024 & 125,784 & 100,921 & 495,631 \\
\hline
7864320 & 7,982 & 15,837 & 31,672 & 62,941 & 125,793 & 118,578 & 475,919 \\
\hline
15728640 & 7,913 & 15,808 & 31,536 & 63,008 & 125,728 & 250,092 & 498,543 \\
\hline
31457280 & 7,930 & 15,883 & 31,582 & 55,060 & 125,540 & 125,300 & 499,833 \\
\hline
\end{tabular}
\end{table}

From the Table \ref{Tab:tab1}, we can observe that for a problem size 983040 the run time on 8 procesors equeals 22,24 sec., then when 32 procesors are used and problem size is increased to 1966080, the run time is very close 23,13 sec. Similarly for 128 processors and problem size equeals 3932160 the run time is 22,07 sec. Therefore we can conclude that the speedup is scaling from 7,94 to 125,78  for workload from 983040 to 3932160 when 128 instead 8 processors are available. 

\section{Conclusions and future work}

In the paper the new way of scalability evaluation of parallel application is proposed. Utilizing the separate measurements of wall-clock time and CPU time, it offers the possibility to estimate the application speedup and efficiency using only the measurement for a single, parallel execution. For the method to be successful it requires only the readily available data, without the need of installation of additional software or application modifications. The experiments performed proved that the estimation accuracy is sensitive to the simplifying assumption taken. For all analysed algorithms the results obtained are similar: the shape of diagrams is similar and the value of speedup is close. In the future works a broader class of algorithms will be taken into consideration, as well as improving the way of weak scalability evaluation will be considered.

\section*{Acknowledgments}

Calculations have been carried out using resources providing by Wroclaw Centre for Networking and Supercomputing (http”//wcss.pl), grant No 266.

\end{document}